\documentclass[journal]{IEEEtran}
\usepackage{cite}
\usepackage{amsmath,amssymb,amsfonts}
\usepackage{algorithmic}
\usepackage{graphicx}
\usepackage{textcomp}
\usepackage{url}
\usepackage{bm}
\usepackage{enumerate}
\usepackage{xcolor}
\usepackage{hhline}
\usepackage{multirow}
\usepackage{bbm}
\usepackage{soul}
\def\BibTeX{{\rm B\kern-.05em{\sc i\kern-.025em b}\kern-.08em
    T\kern-.1667em\lower.7ex\hbox{E}\kern-.125emX}}

\begin{document}
	
\title{Generation of Synthetic Multi-Resolution Time Series Load Data\\}
\author{\IEEEauthorblockN{Andrea Pinceti,
			Lalitha Sankar, and
			Oliver Kosut\\}
\IEEEauthorblockA{School of Electrical, Computer and Energy Engineering\\
Arizona State University\\
}

\thanks{
	
	This material is based upon work supported by the National Science Foundation under Grant No. OAC-1934766 and the Power System Engineering Research Center (PSERC) under projects S-87. 
	
	The authors are with the School of Electrical, Computer and Energy Engineering at Arizona State University, Tempe, AZ, USA.
	
	}

}
\maketitle
	
\begin{abstract}
The availability of large datasets is crucial for the development of new power system applications and tools; unfortunately, very few are publicly and freely available. We designed an end-to-end generative framework for the creation of synthetic bus-level time-series load data for transmission networks. The model is trained on a real dataset of over 70 Terabytes of synchrophasor measurements spanning multiple years. Leveraging a combination of principal component analysis and conditional generative adversarial network models, the scheme we developed allows for the generation of data at varying sampling rates (up to a maximum of 30 samples per second) and ranging in length from seconds to years. The generative models are tested extensively to verify that they correctly capture the diverse characteristics of real loads. Finally, we develop an open-source tool called LoadGAN which gives researchers access to the fully trained generative models via a graphical interface.

\end{abstract}

	
\section{Introduction}\label{intro}

Traditionally, power system research has been based on the development of simplified physical models and the study of their interactions within a system. As a result, a large number of software tools now exists to perform virtually any type of power system analysis \cite{software}. However, the field of machine learning (ML) is experiencing a rapid growth resulting in increased applicability to many fields. For this reason, a large part of the research work within power systems currently focuses on the application of novel ML algorithms and data-driven analysis tools. This change in direction represents a major shift: moving away from physics-based device models to a data-centric analysis of system behaviors. 

Unfortunately, one of the recognized issues which is hindering this process is the lack of widely available datasets of real power system measurements and data. In \cite{sharingdata}, the authors present a review of the current data sharing policies within the electric industry, highlighting several severe limitations as well as the value of increased collaboration. Similarly, the authors of \cite{PFENNINGER2017211} argue that power system data needs to be open and accessible in order to facilitate higher quality research and more effective policy solutions to tackle the challenges of an evolving grid. Moreover, they show that in terms of openness, the energy sector lags behind other industries. Another critical aspect is the fact that open availability of data and models reduces the doubling of work and efforts \cite{MEDJROUBI201714,PFENNINGER2017211} and allows to more effectively build on previous research. 

While synthetic grid models are increasingly being used as the basis for many research projects \cite{birchfield,birchfiled2,birchfield3,eran}, one of the drawbacks that limits their usefulness is the fact that they often only provide a fixed operating point. Most power system studies as well as any machine learning-based application rely on the availability of large datasets of measurements to capture the complex behaviors of electrical systems. For this reason, the ability to map time-series data to a specific grid model is crucial to enable the development of new and improved applications. 

We developed an open source tool for the generation of synthetic time-series bus-level load data at varying sampling rates and for different time lengths. We focus on the time-varying characteristics of system loads as they represent one of the major variables driving power system behaviors and depend on phenomena which is external to the electrical system itself (consumer habits, weather, etc.). Moreover, while load data is used for virtually every power system study, the requirements in terms of sampling rate and time length vary based on the specific application. Thus, our objective is to create a complete and universal tool which can generate synthetic load data at any scale, starting from the available PMU speed of 30 samples per second. 

Different methods for the generation of time-series load data have been previously proposed. For example, \cite{hanyue} presents a bottom-up approach in which each bus-level load is obtained as a combination of typical load curves describing different types of customers (residential, commercial, etc.). This approach was used to generate time-series load data for the synthetic grid models developed at Texas A\&M \cite{birchfield}. The limitations of this method are twofold: first, it can only be applied to models for which geographical and demographical information is available and second, because it uses a limited number of base profiles, the generated data is limited in diversity and variability. These limitations can be overcome by adopting data-driven approaches which leverage large amounts of real data. Examples of these have been proposed for the generation of renewable scenarios \cite{baosen}, consumer level smart-meter data \cite{smartmeterdata,consumer,en13010130,LAN20181188} and net load time-series data \cite{zonal1}. However, these learning approaches have never been used for bus-level loads at the transmission level. Moreover, these existing techniques are limited in scope since they can only generate data for limited time-horizons and at a fixed sampling rate. In contrast, our proposed generative scheme addresses a significantly larger scope, both in terms of theoretical contributions, as well as range of applicability. 

\begin{figure}
	\centering
	\includegraphics[scale=0.38]{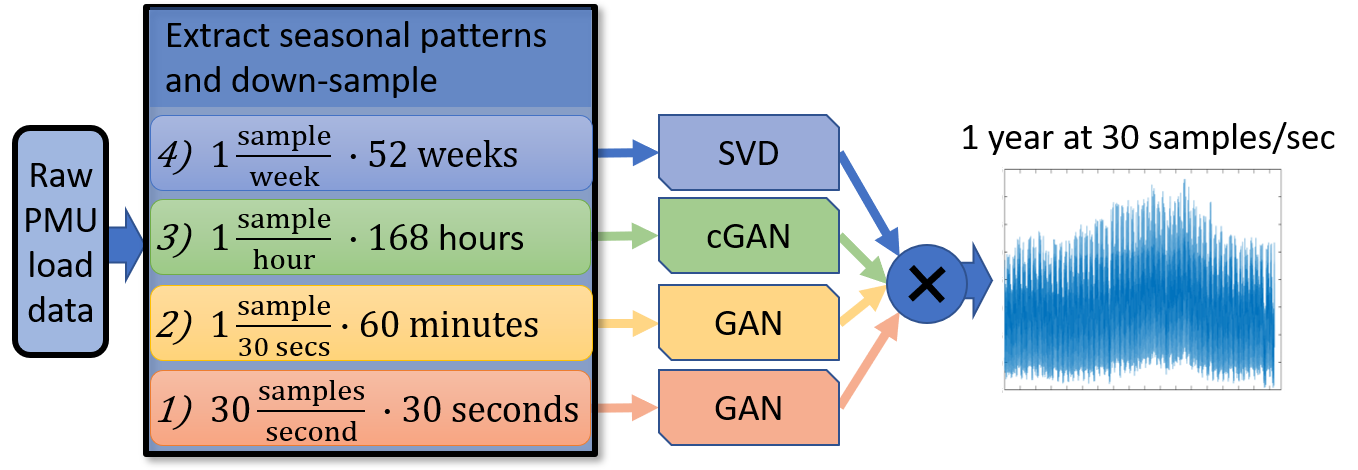}
	\caption[Overview of the Generative Scheme.]{Overview of the complete generative scheme illustrating the four down-sampling levels and their respective generative models.}
	\label{overview} 
\end{figure}

\subsection*{Contributions}

We designed a complete generative framework which can create any number of unique bus-level load profiles at sampling resolutions from 30 Hz to one sample per year and for any time length ranging from a few seconds to a year. We leverage a proprietary dataset of over 70 Terabytes of phasor measurements to learn and model the behavior of system loads from sub-second to yearly time-scales. Based on the different temporal characteristics of the loads, two main types of generative models are used: generative adversarial networks (GANs) and principal component analysis (PCA). As described in Section~\ref{loadmodelling}, the proposed scheme relies largely on the use of GANs since it has been shown that this type of powerful, nonlinear methods are far more effective in generating synthetic datasets than simpler linear models \cite{Mei2018}. The inadequacy of linear models to capture the richness of the load dataset used in this work is illustrated experimentally in Section~\ref{validation}.
Via a process of data down-sampling and aggregation at different levels, we train independent generative models to capture the characteristics of distinct load profiles. This process can be summarized in four steps:
\begin{enumerate}[(I)]
	\item The raw PMU data (voltages and currents) is used to compute power consumption at different load buses;
	\item The PMU-speed load data is down-sampled at four different levels:
	\begin{enumerate}[1)]
	    \item 30-second-long profiles at 30 samples/second
	    \item hour-long profiles at 1 sample/30 seconds
	    \item week-long profiles at 1 sample/hour
	    \item year-long profiles at 1 sample/week
	\end{enumerate}
	
	\item A generative model is trained individually for each level;
	\item The synthetic data generated using the trained models is combined to obtain any desired time-length and sampling resolution.
\end{enumerate}

Figure~\ref{overview} shows an overview of the complete generative scheme. The four aggregation levels have been chosen as a compromise between having profiles of comparable size at each level while dividing the data in such a way that each level captures different and independent behaviors. In fact, the first and second levels from the bottom model the fast time-scale load variations, the third level captures the daily load changes, while the fourth level captures yearly seasonality. Nonetheless, regardless of the choice of aggregation levels, the final generative scheme can create data at any arbitrary resolution from the highest limit of 30 samples/second (the raw PMU sampling rate) to any lower sampling rate. 

The remainder of the paper is organized as follows. In Section~\ref{over} an overview of GANs and the dataset used is presented. Section~\ref{loadmodelling} describes the details of the generative models used and how they are trained. The process of combining the generative models to obtain specific sampling resolutions and time lengths is shown in Section~\ref{combining}, while Section~\ref{practical} describes the design of the open-source generative tool and its validation. Supporting documents, including the Appendix, source code, and additional figures, can be found at \cite{appendix}.

\section{Overview}\label{over}

\subsection{Conditional Generative Adversarial Networks}\label{cgan}
The proposed generative framework builds upon previous results on the generation of synthetic time-series load data. In particular, it relies on the use of GANs \cite{goodfellow2014generative} and conditional GANs (cGANs) \cite{mirza2014conditional}. This machine learning-based class of generative models consists of two networks, a discriminator and a generator, competing against each other with the goal of iteratively improving their performance. In particular, the discriminator is trained to distinguish between real data and synthetically generated data, while the generator aims at creating synthetic data which cannot be discerned from the real one. A cGAN represents a modification of this basic framework, where the generator is trained to learn the conditional distribution of the data. By using conditioning labels to describe specific characteristics of the real data, the generation of the synthetic data can be controlled and targeted towards specific classes of data. For example, in the context of time series load data, unique behaviors can be observed based on the characteristics of each load, resulting in different data distributions. Using cGANs allows to model these different characteristics by learning the conditional distribution of the data.

\begin{figure}
	\includegraphics[scale=0.2]{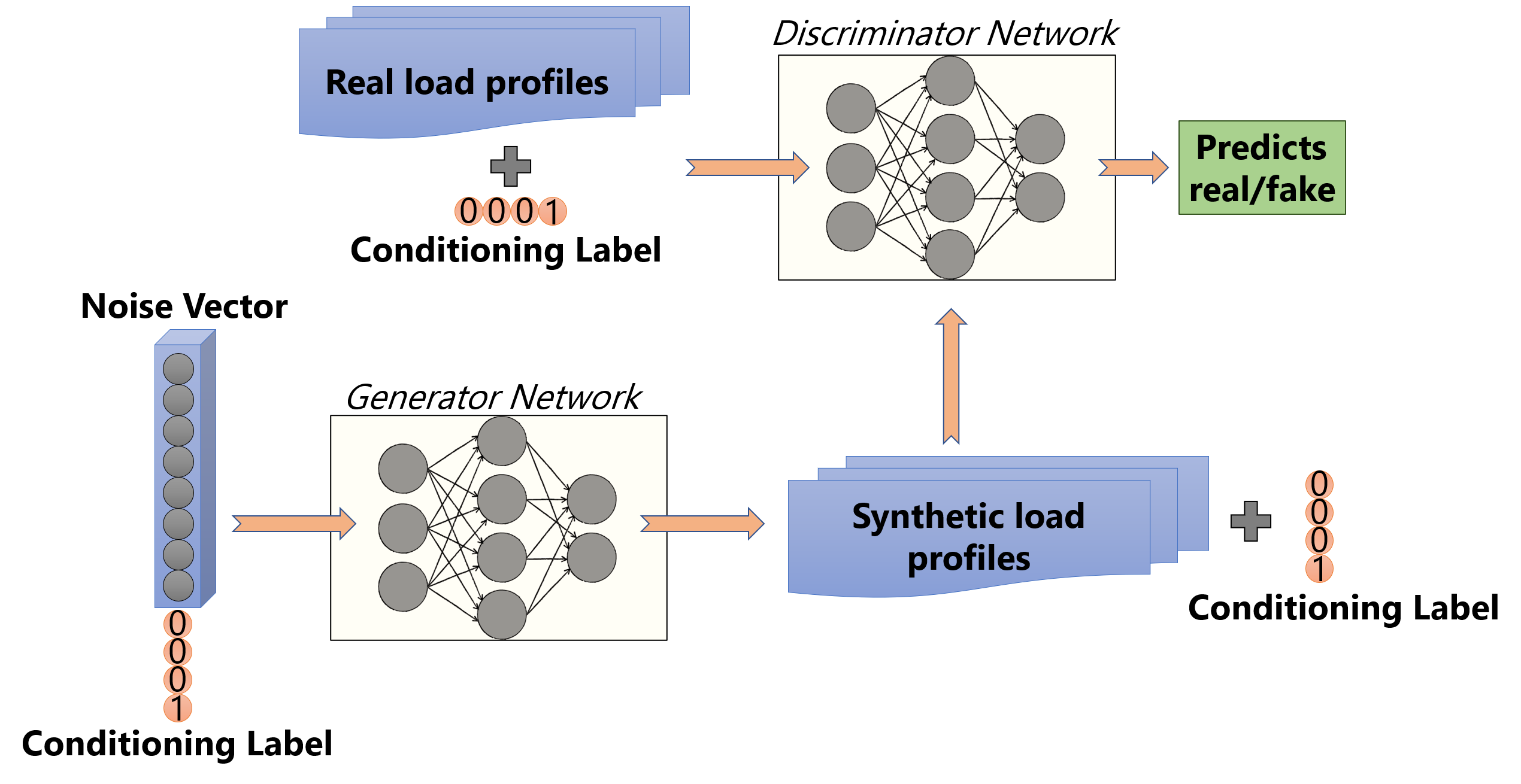}
	\caption{Architecture of a conditional generative adversarial network.}
	\label{condGAN} 
\end{figure}

In \cite{Pinceti2021PESGM}, we showed how a cGAN can be trained to learn the behavior of power system loads and subsequently generate realistic time-series load data conditioned on two labels: the time of the year and the load type. The basic architecture of the cGAN used is shown in Fig.~\ref{condGAN}. The resulting model was tested extensively to verify that the synthetic data produced by the cGAN correctly captured the characteristics of the real load data. In this work, we build upon those preliminary results to develop several generative models which can be combined to create time-series data at varying resolutions and varying time-lengths.

\subsection{Dataset Description}
The goal of the generative scheme is to create synthetic load data to be used for a large number of diverse downstream applications. To achieve this, the training data must be able to capture load behaviors at every time resolution to provide a complete assessment of the characteristics of different system loads. In this work, we leverage a large dataset of real phasor measurement data (upwards of 70 Terabytes) obtained from a major US electric utility. This data was collected for two consecutive years from almost 500 phasor measurement units (PMUs). Based on the system topology and placement of the PMUs, 12 fully monitored load buses were identified. Using the complex current and voltage measurements on each line, the net power injection (corresponding to the load demand at that bus) was computed. The availability of time-series load data measured at 30 samples per second for multiple consecutive years allows for the observation of load behaviors at different time-scales. 

\begin{figure}
	\centering
	\includegraphics[scale=0.33]{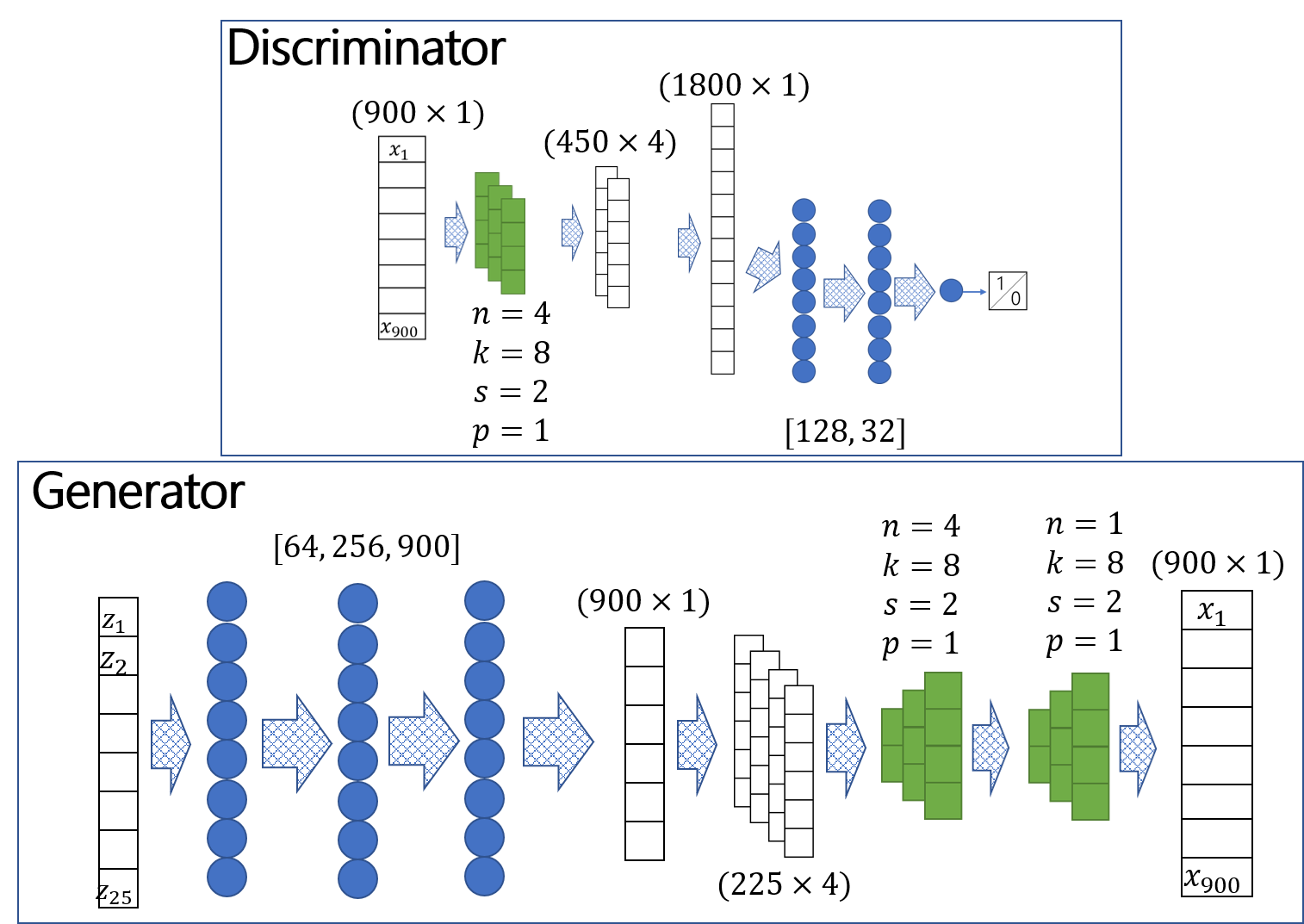}
	\caption{Architecture of the GAN for the 30-seconds-long profiles.}
	\label{30secGAN} 
\end{figure}

Given its size and extent, the dataset used captures several distinct types of load behaviors which need to be correctly modelled by the generative scheme to create realistic and diverse synthetic data. Two main factors determine the characteristics of the load profiles: the type of load and the time of the year. Based on the percentage composition of each load in terms of residential, commercial, and industrial components, the loads have been grouped into two classes: mainly industrial loads and mainly residential loads. This classification is the result of the fact that loads present very distinct behaviors depending on their composition; it was observed that loads where the main component was either residential or commercial showed very similar behaviors, with the demand varying very smoothly and regularly. On the other hand, mainly industrial loads behaved more unpredictably resulting in irregular profiles. A similar difference is observed based on the time of the year; for example, the daily consumption during summer follows a very regular and smooth curve, while in winter the loads fluctuate more and present several peaks during the day. As described in the next section, depending on the time-scale under study, these elements have varying influence on the observed load behaviors. Nonetheless, it is important that the generative models used are able to capture these differences in load characteristics in order to generate realistic data for each type of load.

\section{Load Modelling}\label{loadmodelling}
In this section, the down-sampling of the time-series data and the generative models used are described for each level. 

\subsection{Level 1 --- 30-second-long profiles}
At the lowest level, the load profiles are chosen to be 30 seconds long and sampled at the PMU rate of 30 samples per second. In order to be able to feed this data to the learning algorithm, each profile is individually normalized by dividing it by its average load value.

\begin{figure}
	\centering
	\includegraphics[scale=0.3]{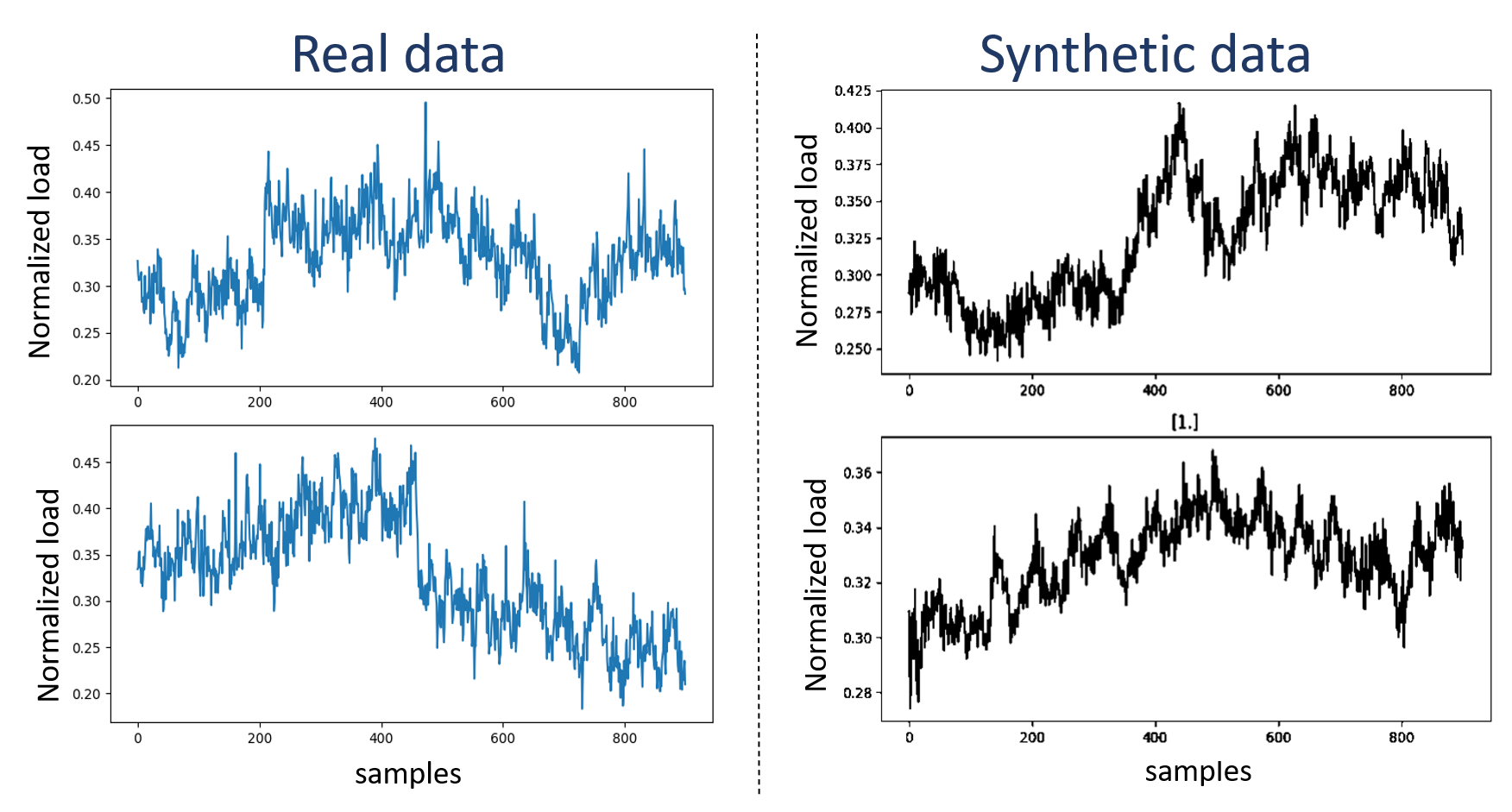}
	\caption{Comparison between two real 30-second-long profiles (left) and two synthetic 30-second-long profiles (right).}
	\label{30sec_comp} 
\end{figure} 

The behavior of the 30-seconds-long profiles is learned and modelled using a basic GAN as described in Section \ref{cgan}. Given the temporal nature of the load data, the generator and discriminator are both represented by 1-dimensional (1D) convolutional neural networks of similar complexity. The discriminator uses a 1D convolutional layer to extract a spectral representation of the input profile and two fully-connected layers to decide if the profile is real or synthetic. Similarly, the generator takes the random vector of noise as input and it passes first through three fully-connected layers and then through two 1D transpose convolutional layers which perform an upscaling to the final output profile. The detailed architecture of the GAN used is shown in Fig.~\ref{30secGAN}.

In order to ensure a robust and successful training of the GAN, the generator and discriminator are trained following an iterative process. For each batch in an epoch, the discriminator is trained twice for each update of the generator. This gives the discriminator a slight edge, in turn forcing the generator to improve the quality of the synthetic profiles. For a visual representation of the results of the training process, Fig.~\ref{30sec_comp} shows a comparison between two real and two synthetically generated 30-second-long profiles. It is to be noted that while the generated data captures the behavior of the real data, the generator does not simply copy real profiles but instead creates completely new samples. 

To quantitatively verify that the synthetic data retains the characteristics of real loads, two statistical tests are performed. First, the Wasserstein distance, a metric that measures the distance (or similarity) between two distributions, is used to compare the real and synthetic data. Given the set of all possible joint distributions $\Pi(p_r,p_g)$ over distributions $p_r$ and $p_g$, the Wasserstein distance $W(p_r,p_g)$ between them is defined as
\begin{equation}
    W(p_r,p_g) = \inf_{\gamma \sim \Pi(p_r,p_g)} \mathbb{E}_{(x,y)}[\|x-y\|]
\end{equation}

\begin{figure}
	\centering
	\includegraphics[scale=0.8]{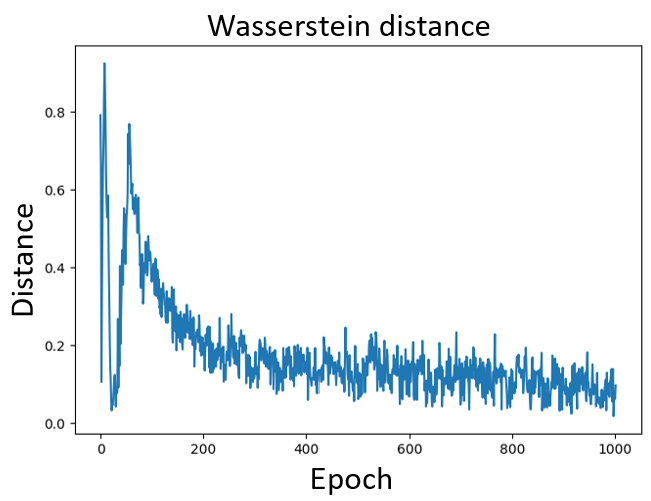}
	\caption{Wasserstein distance between real and generated 30-seconds-long data as a function of the epochs.}
	\label{30sec_wd} 
\end{figure}

In practice, we computed the Wasserstein distance over discrete distributions estimated by obtaining the histograms of the real and generated datasets over a randomly selected subset of 500 samples each. The distance is computed at each epoch and it is shown in Fig.~\ref{30sec_wd} as a function of the epoch number: at the beginning, the distance is relatively high, while as training progresses the distance drops and approaches zero. This means that the generator is correctly capturing the distribution of the real data. The second metric used is the power spectral density (PSD), which is defined as
\begin{equation}
    S(\omega)=\frac{1}{2\pi}\int_{-\infty}^{\infty}dt ~r(t)e^{-i\omega t}
\end{equation}
where $r(t)$ is the autocorrelation function for a given signal $f(t)$
\begin{equation}
    r(t)=\lim_{T\to\infty}\frac{1}{2T}\int_{-T}^{T} d\tau~ f(\tau)f^*(\tau+t)
\end{equation}

\begin{figure}
	\centering
	\includegraphics[scale=0.65]{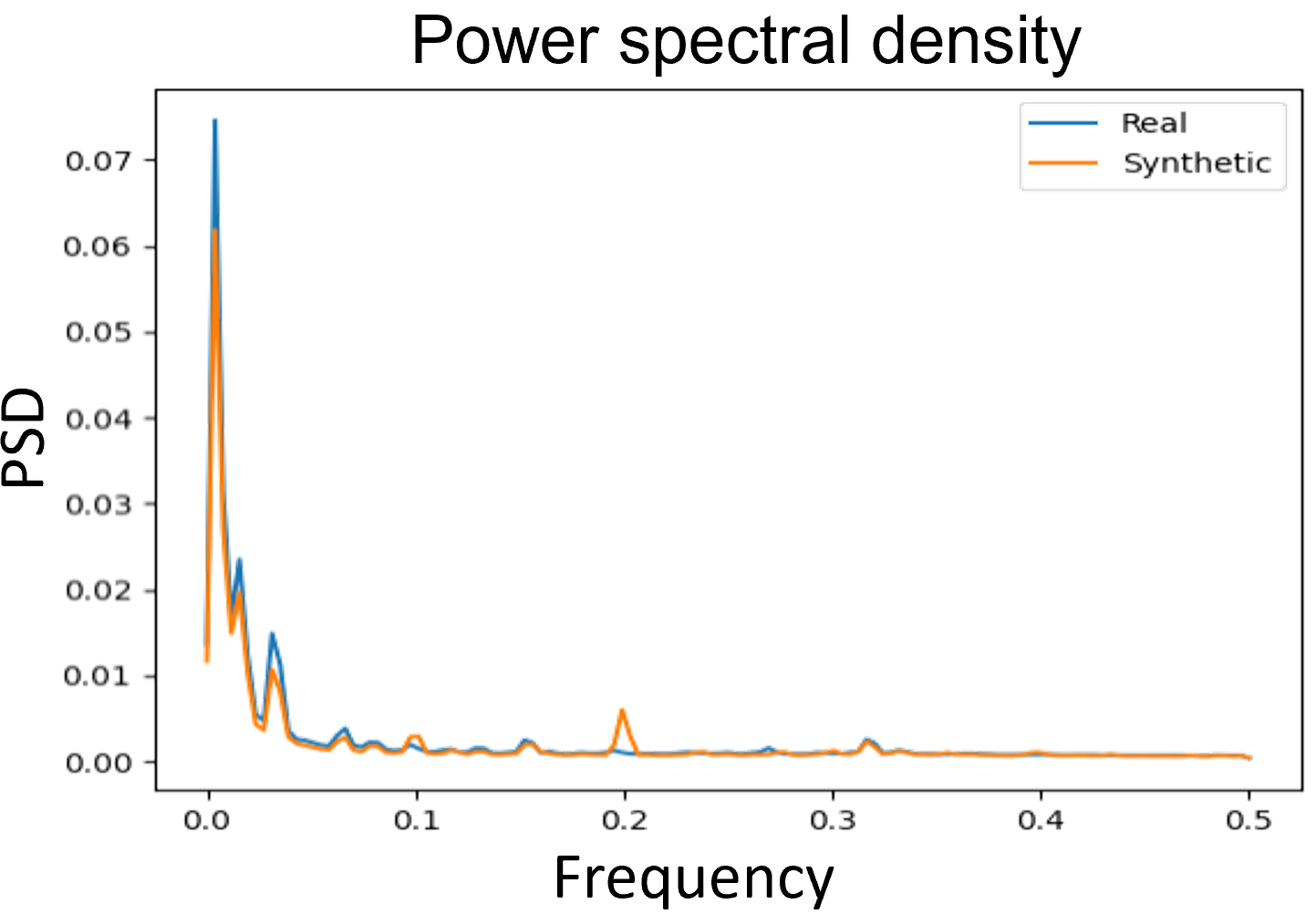}
	\caption{Power spectral density comparison between real (blue) and synthetic (orange) 30-seconds-long datasets.}
	\label{30sec_psd} 
\end{figure}

The PSD allows us to observe the temporal characteristics of the profiles. Figure~\ref{30sec_psd} shows a comparison between the PSD of real and synthetic data, verifying that they both present the same overall behavior.

\subsection{Level 2 --- Hour-long profiles}\label{hour_long}
\begin{figure}
	\centering
	\includegraphics[scale=0.5]{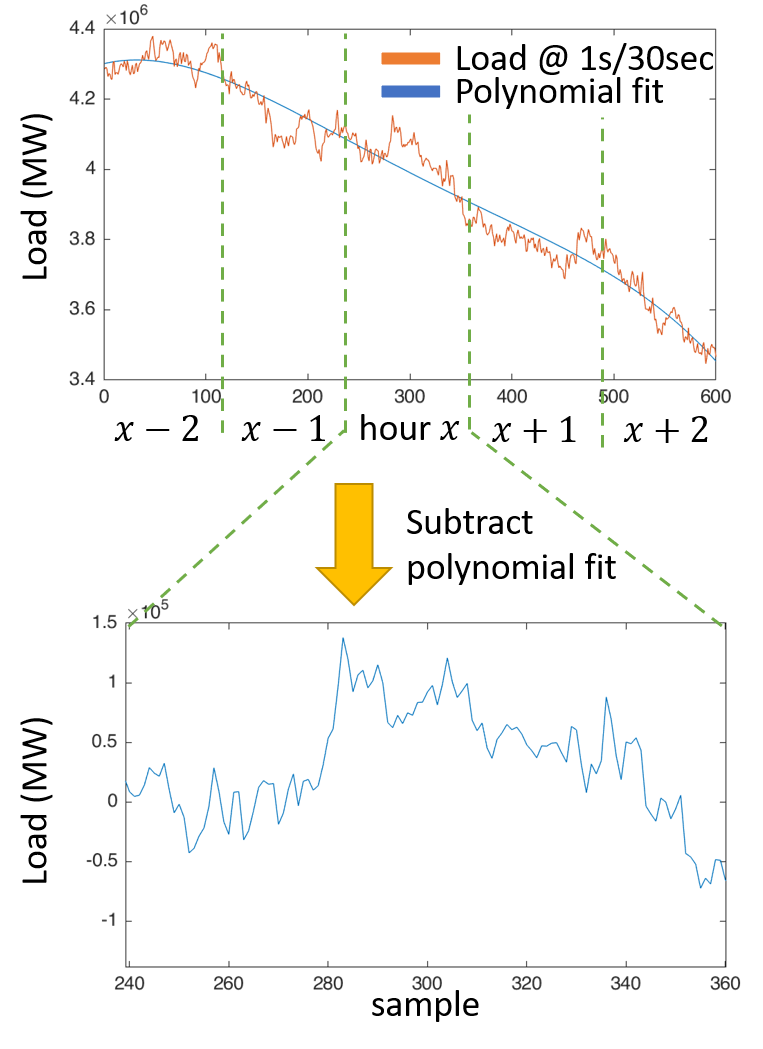}
	\caption{Example of the de-trending process for hour-long profiles. A polynomial line is fitted to 5 hours of data (top) and then subtracted from the hour to be normalized (bottom).}
	\label{hour_detrend} 
\end{figure}

The profiles at this level are sampled at 1 sample every 30 seconds for a length of one hour; this results in each sequence being 120 samples long. Each point is computed as the average load over the 30 second period. The normalization of each profile is performed by dividing by the average load over the one hour period and by de-trending the resulting sequence. The de-trending step is necessary because, at this resolution and time length, the load data presents two main patterns: slow, large-amplitude trends and smaller, faster load variations. The slower trends represent the overall change in load consumption from one hour to the next; these are fully captured by the hourly data from the third level and for this reason they do not need to be modeled here. By de-trending the signals, we extract the faster time-scale load changes which happen at a sub-minute scale. For each hour-long profile, the de-trending is performed by fitting a $4^{th}$ degree polynomial to the data from two hours before the selected hour to two hour after. Then, the fitted curve is subtracted from the hour-long profile. Figure~\ref{hour_detrend} illustrates the polynomial fitting and de-trending steps. This process allows to better capture the real trend by taking into consideration the behavior of the load before and after the hour to be normalized. As shown in the next section, this guarantees a more realistic output when generating large amounts of synthetic data. 

\begin{figure}
	\centering
	\includegraphics[scale=0.3]{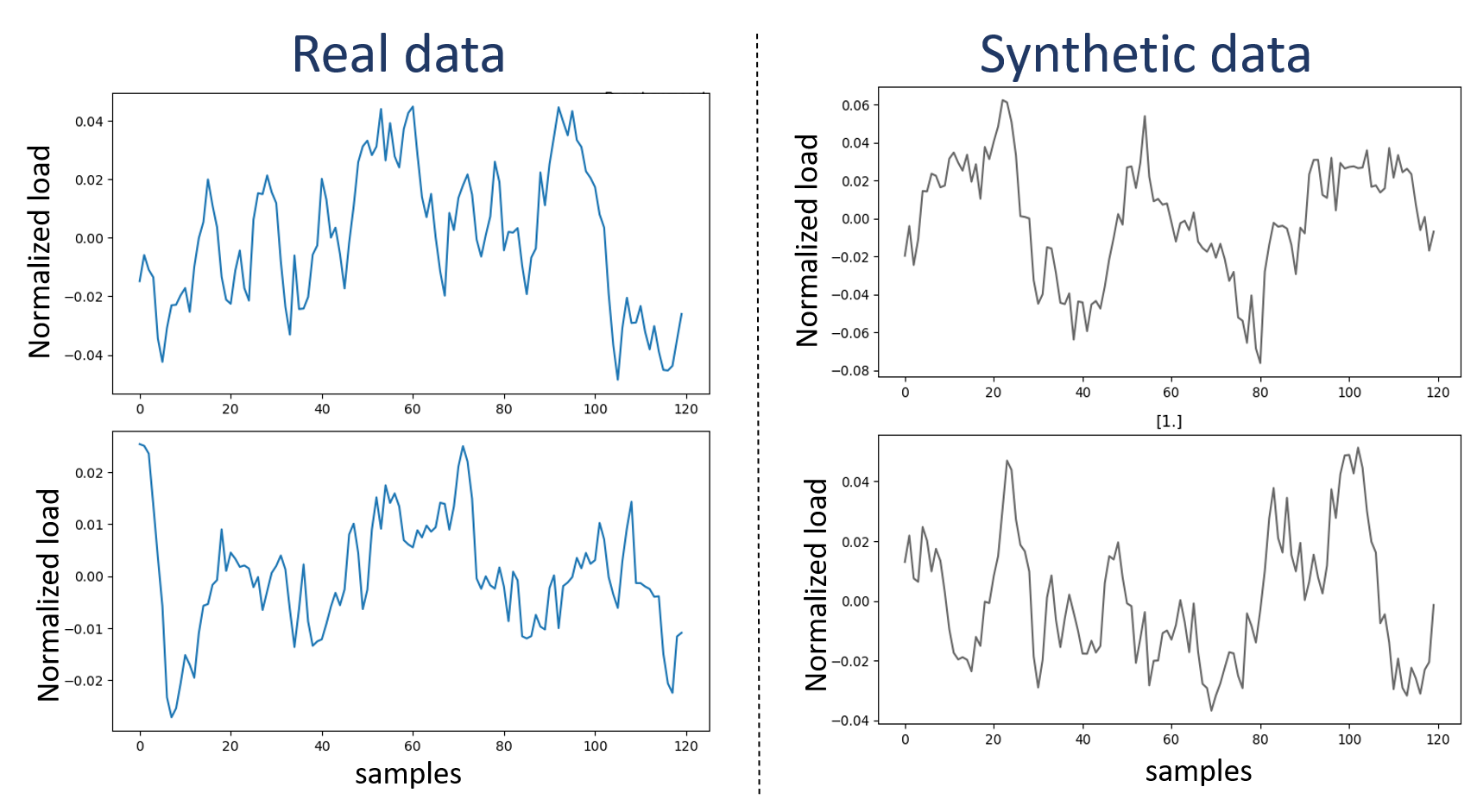}
	\caption{Comparison between two real hour-long profiles (left) and two synthetic hour-long profiles (right).}
	\label{hour_comp} 
\end{figure}

The profiles at this level are modelled using a similar GAN to the one described in the previous section. Most of the hyper-parameters (such as those related to the loss function, learning rate and decay rate) are directly transferred from the 30-seconds-long GAN and only required minimum adjusting. The same training strategy is also used here: the discriminator is updated twice as often as the generator in order to achieve a more robust training convergence. 

The synthetic hour-long data resulting from the trained GAN is validated following the same procedure as described for the 30-seconds-long profiles in the previous section. The detailed results as well as the GAN architecture can be found in the Appendix \cite{appendix}. For a visual comparison, Fig.~\ref{hour_comp} shows randomly selected examples of real and synthetic hour-long profiles.

\subsection{Level 3 --- Week-long profiles}
At this level, the time-series load profiles are computed to be one week long and sampled at 1 sample per hour. Each sample represents the average load over the corresponding one hour period. Each week-long profile is also normalized by dividing it by its corresponding average weekly load. 

To model the load profiles at this level, we use the cGAN we developed in \cite{Pinceti2021PESGM}. The generation of week-long synthetic data is conditioned on two labels simultaneously: the load type (mainly residential or mainly industrial) and the season. In \cite{Pinceti2021PESGM}, we showed that the 1-dimensional convolutional neural networks chosen for the generator and the discriminator allowed for the successful training of the cGAN. The quality of the resulting synthetic data was verified based on the statistical tests used in the previous two sections as well as downstream applications. The synthetic data was successfully used to train a load forecasting algorithm and to perform AC optimal power flow on a test system. The complete validation results for the cGAN can be found in \cite{Pinceti2021PESGM} as well as in the Appendix \cite{appendix}.

\subsection{Level 4 --- Year-long profiles}
The year-long profiles from the top level, which have a resolution of 1 sample per week, capture the long-term monthly and seasonal patterns. For each load, one year of data is down-sampled by computing the average weekly load, resulting in a profile 52 samples long. Since 12 loads are available over two years, a total of 24 year-long profiles are obtained. At this time-scale and resolution, residential and industrial loads present very different behaviors: the former show distinct peaks during summer and winter, while the latter are more constant throughout the year. 


Given the limited diversity and complexity of the load profiles at this level, along with the small number of real profiles available, the generative model presented in \cite{pincetiPESGM19} is used. This approach is based on using singular value decomposition (SVD) as a way to extract prototypical load patterns from the real data and subsequently generate new, realistic data by taking linear combinations of them. This model, while simpler than the GANs used for the other levels, is sufficient to capture the main behaviors of these loads and it should be preferred given its higher interpretability. In fact, \cite{GAN_PCA} shows that methods based on principal component analysis (such as the SVD-based approach used here), under certain conditions, can fully describe Gaussian data. Given the highly averaged nature of the load profiles at this level, it is reasonable to assume for this to apply to the year-long time-series data.

\begin{figure}
	\centering
	\includegraphics[scale=0.38]{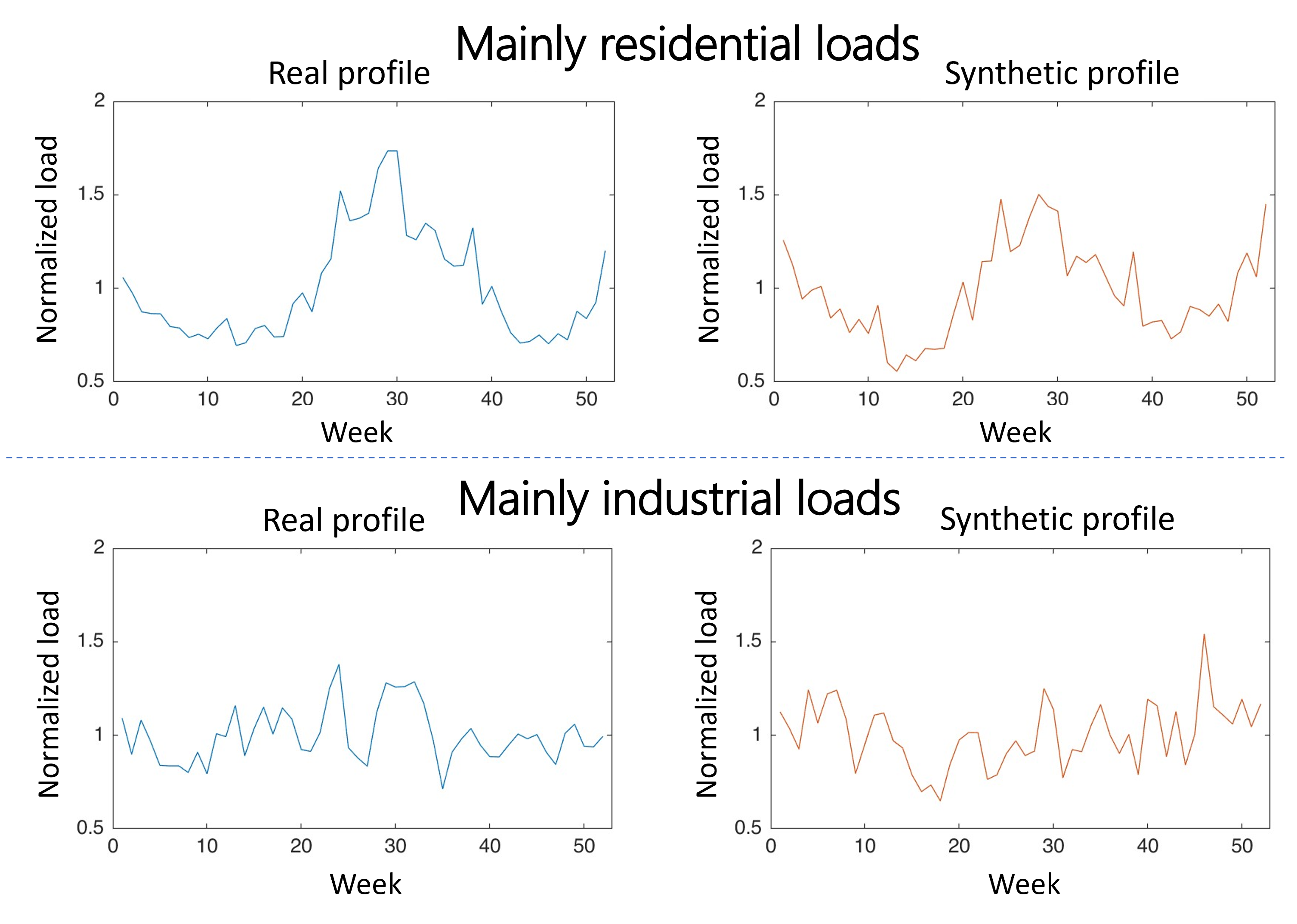}
	\caption{Comparison between real (left) and generated (right) year-long profiles for mainly residential loads (top) and mainly industrial loads (bottom).}
	\label{year_rescomp} 
\end{figure}

Two independent models are learned for the two types of loads (mainly residential and mainly industrial). Each dataset has size of $12 \times 52$ (12 profiles, each with 52 samples). Given a load matrix $L \in \mathbb{R}^{12\times52}$, singular value decomposition is performed as follows:
\begin{equation}
L=U \Sigma V^T
\end{equation}
where the rows of $V^T$ correspond to the prototypical load patterns and the singular values in the diagonal matrix $\Sigma$ represent their respective weights. Thus, each load profile is a linear combination of the weighted yearly patterns based on their unique coefficients in each row of the $U$ matrix. To generate new, realistic profiles, first the distributions of each of the 12 columns of the coefficient matrix $U$ are learned. Then, a new coefficient vector is sampled and multiplied by $\Sigma V^T$. This process is performed independently for mainly residential and mainly industrial loads. Figure~\ref{year_rescomp} shows a comparison between real and generated profiles for the two load types. 


\section{Multi-resolution synthetic data}\label{combining}
Using the trained generative models for each aggregation level, it is possible to generate synthetic data for any time-length and any time-resolution by combining multiple profiles together. In particular, to obtain longer times, several profiles at a given level are concatenated together. Furthermore, to obtain a specific resolution, profiles at different levels can be combined together. In Sections~\ref{horcomb} and \ref{vercomb} these two types of aggregation are described and some examples are shown. Section~\ref{anyres} presents a method to obtain any specific user-defined time-resolution without being limited to the predefined resolutions of the four original aggregation levels. 

\subsection{Concatenating load profiles}\label{horcomb}
At each level, a trained generative model creates profiles of fixed length, based on the length of the real profiles used for training. To obtain a longer time duration, multiple profiles from a given level need to be concatenated together. For example, if four weeks of data are required, the cGAN for Level 3 is used to generate four individual week-long profiles which are then concatenated together to obtain the final time-length. Depending on the specific scenario, attention needs to be paid to ensure that the correct load characteristics are maintained; for example, the load type and season labels must be consistent when generating the four week-long profiles.

While this approach is very straightforward and it naturally follows from the load aggregation structure chosen, it presents one challenge: ensuring continuity at the seam between two concatenated profiles. Because consecutive profiles are generated independently, concatenating them could result in some degree of mismatch between the end of one profile and the start of the next one. 

For the profiles at Level 3 (week-long profiles), a linear filter is used to smoothen the seam between two profiles. The linear filter is applied to the last two points of one profile and the first two points of the following one. Moreover, the filter is chosen to have a size of five; that means that each point is adjusted as a linear combination of itself and the previous two and following two values, that is
\begin{equation}
	x_{k}^{\text{new}} = \sum_{i=-2}^{2} \beta_{i} x_{k+i}
\end{equation}
where, $x_k$ is the sample to which the filter is applied, $x_{k}^{\text{new}}$ is the new, filtered value and $\beta$ is the vector of filter weights. The specific values of the filter are learned from the real load dataset by solving the following least-square problem:
\begin{equation}
	X_{[-2,-1,1,2]}\ \beta_{[-2,-1,1,2]} = 0.5X_k
\end{equation}
where, $X_{[-2,-1,1,2]}$ is a matrix in which each row represents the two previous values and two following values of an entry $x_k$ and $\beta_{[-2,-1,1,2]}$ are the corresponding filter values. $X_k$ is the column vector of all values $x_k$ and using the factor 0.5 corresponds to fixing the value of $\beta_0 = 0.5$. 



\begin{figure}
	\centering
	\includegraphics[scale=0.49]{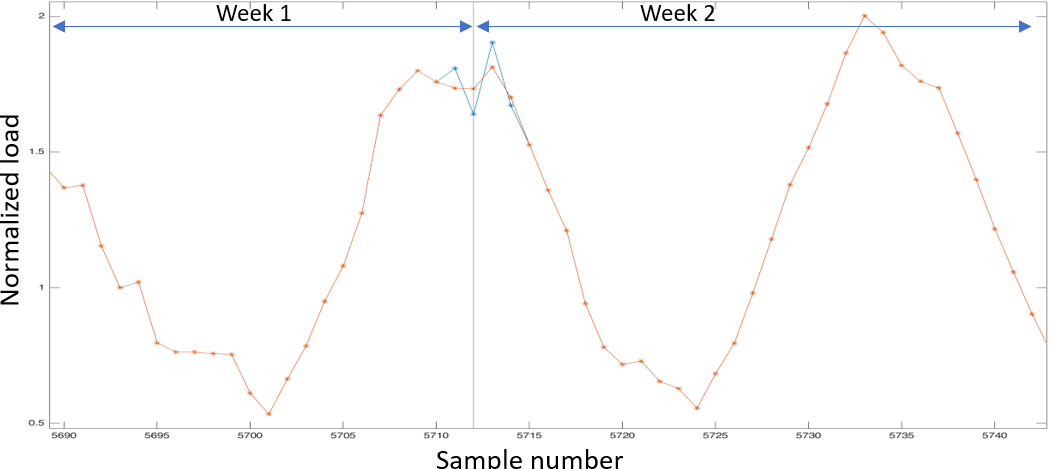}
	\caption{Worst-case example of discontinuity between two concatenated summer profiles. The blue curve shows the load values before the filter is applied and the orange curve shows the filtered values which eliminate the discontinuity. The vertical line indicates the time coordinate of the last sample of the previous profile.}
	\label{filter_summer} 
\end{figure} 

Figure~\ref{filter_summer} shows an example of the discontinuity that can be observed when concatenating two profiles from Level 3. The blue points represent the load values before applying the filter; the red curve shows the values after the linear filter is used. It can be seen that without the filter there is a sharp discontinuity at the seam between the two weeks; applying the filter results in a more regular load behavior. It is to be noted that the example shown represents a worst case scenarios, where the discontinuity is relatively large; in most cases, the concatenation does not result in a detectable discontinuity.  

\begin{figure}
	\centering
	\includegraphics[scale=0.40]{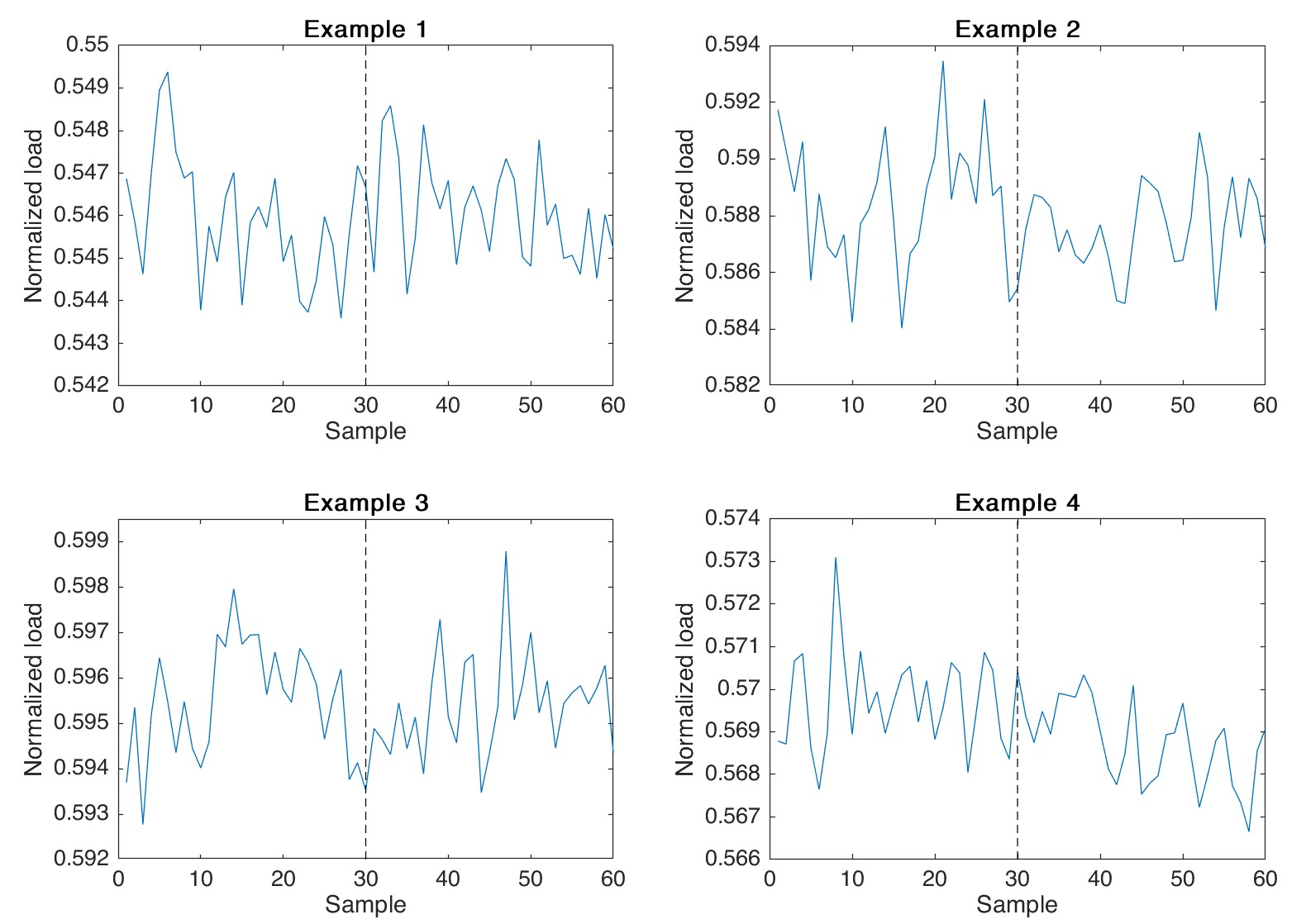}
	\caption{Examples of concatenation of 30-second-long profiles from Level 1. Each plot shows two seconds of data at 30 sample/second obtained by concatenating two separate profiles. The first 30 samples represent the end of one profile and the second 30 samples are the beginning of the following profile. The vertical line indicates the last sample of the previous profile. In all cases, it can be seen that the concatenation does not produce any recognizable discontinuity.}
	\label{30sec_concat_examples} 
\end{figure}

At Levels 1 and 2, it was observed that concatenating profiles does not yield the same discontinuity issues as observed for Level 3. This behavior is to be expected because at these time scales the load fluctuations are much smaller compared to the hourly changes in the week-long profiles and the behavior of the loads is less regular and predictable. For these reasons, the load change between two consecutive profiles at Level 1 or 2 is indistinguishable from the normal load behavior. To better illustrate this, Fig.~\ref{30sec_concat_examples} shows four separate examples of concatenation of profiles from Level 1. Each plot shows two seconds of data at 30 sample/second obtained by concatenating two separate profiles. The first 30 samples represent the end of one profile and the second 30 samples are the beginning of the following profile. In all cases, it can be seen that the concatenation does not produce any recognizable discontinuity. While visual inspection of several generated profiles provided a qualitative confirmation of this phenomena, a statistical analysis is performed to further verify this observation. In particular, the percentage difference between the last sample of a profile and the first sample of the following profile is measured for a dataset of real data and a dataset of synthetically generated data. Table~\ref{concat_analysis} shows a comparison of the mean and standard deviation of the percentage difference observed in the real and synthetic datasets, for both profiles at Level 1 and Level 2. The results of this statistical analysis confirm that a linear filter is not required to concatenate profiles from the two bottom levels since no significant difference can be observed between the seams of real profiles and generated profiles.

\begin{table}[]
	\centering
	\caption[Comparison of the Percentage Difference Between Consecutive Samples in Real and Generated Data, for Level 1 and Level 2.]{Comparison of the percentage difference between consecutive samples in real and generated data, for Level 1 and Level 2.}
	\label{concat_analysis}
	\begin{tabular}{|c|c|c|c|}
		\hline
		\multicolumn{1}{|l|}{\multirow{2}{*}{\textbf{Aggregation Level}}} & \multicolumn{1}{l|}{\multirow{2}{*}{\textbf{Dataset}}} & \multicolumn{2}{l|}{\textbf{Percentage Difference}} \\ \cline{3-4} 
		\multicolumn{1}{|l|}{} & \multicolumn{1}{l|}{} & \begin{tabular}[c]{@{}c@{}}Mean\\ {[}\%{]}\end{tabular} & \begin{tabular}[c]{@{}c@{}}Std. Dev.\\ {[}\%{]}\end{tabular} \\ \hline
		\multirow{2}{*}{Level 2} & Real & 0.59 & 0.49 \\ \cline{2-4} 
		& Synthetic & 0.47 & 0.33 \\ \hline
		\multirow{2}{*}{Level 1} & Real & 0.17 & 0.14 \\ \cline{2-4} 
		& Synthetic & 0.13 & 0.27 \\ \hline
	\end{tabular}
\end{table}

\subsection{Combining profiles at different resolutions}\label{vercomb}
In the previous section, we showed how synthetic load data of any length can be obtained by concatenating multiple profiles. In order to capture behaviors at different time-scales and to generate data at a specific resolution it is necessary to also combine profiles from different levels. For example, if we are interested in hourly load data over an entire year, first 52 week-long profiles must be concatenated (following the correct seasons) to get the desired resolution (1 sample/hour) and secondly, they must be combined with one year-long profile from the fourth level to capture the long-term seasonal behavior of the load. The process of combining different levels follows directly from the aggregation scheme which was adopted initially to down-sample the load profiles at different resolutions. The steps to combine the different levels are as follows.

(i) \textit{Combining Levels 4 and 3.} The profiles from Level 4 (year-long profiles at 1 sample/week) capture the slow and long term changes in energy demand across months and years, while Level 3 (week-long profiles at 1 sample/hour) captures the daily load behavior. The real dataset of week-long profiles was normalized by dividing each profile by their average weekly load; at the same time, each point in the year-long profiles was computed as the average demand during any given week. This means that to combine the two levels together, each week-long profile must be scaled by the corresponding value in the year-long profile. The scaling is performed by multiplying the week-long profile at week $i$, defined as $X_2^i$, by the factor $x_{1,i} / \text{mean}(X_2^i)$, where $x_{1,i}$ is the value corresponding to the $i^{th}$ week from the year-long profile $X_1$. By doing so, the week-long profile is effectively normalized so that its average equals the corresponding value of the year-long profile. 

(ii) \textit{Combining Levels 3 and 2.} For resolutions up to 1 sample every 30 seconds, the profiles from Level 3 need to be combined with profiles from Level 2 (hour-long profiles at 1 sample/30 seconds). The process of combining the hour-long profiles with a week-long profile consists of two steps: 1) scaling the hour-long profiles and 2) adding the polynomial trend. As we showed in Section~\ref{hour_long}, the last step in creating the hour-long profiles consisted in normalizing them by dividing by the average hourly load. For this reason, when generating the synthetic data, we first scale each hour-long profile by the corresponding value in the week-long profile following the same procedure as with Levels 4 and 3 above. Secondly, a polynomial trend is added to each hour-long profile in order to capture the slower load variations that can be observed from the week-long profile. The trend is computed by fitting a $4^{th}$ degree polynomial to the points in the week-long profile, from two hours before the hour of interest to two hours after. 

\begin{figure}
	\centering
	\includegraphics[scale=0.26]{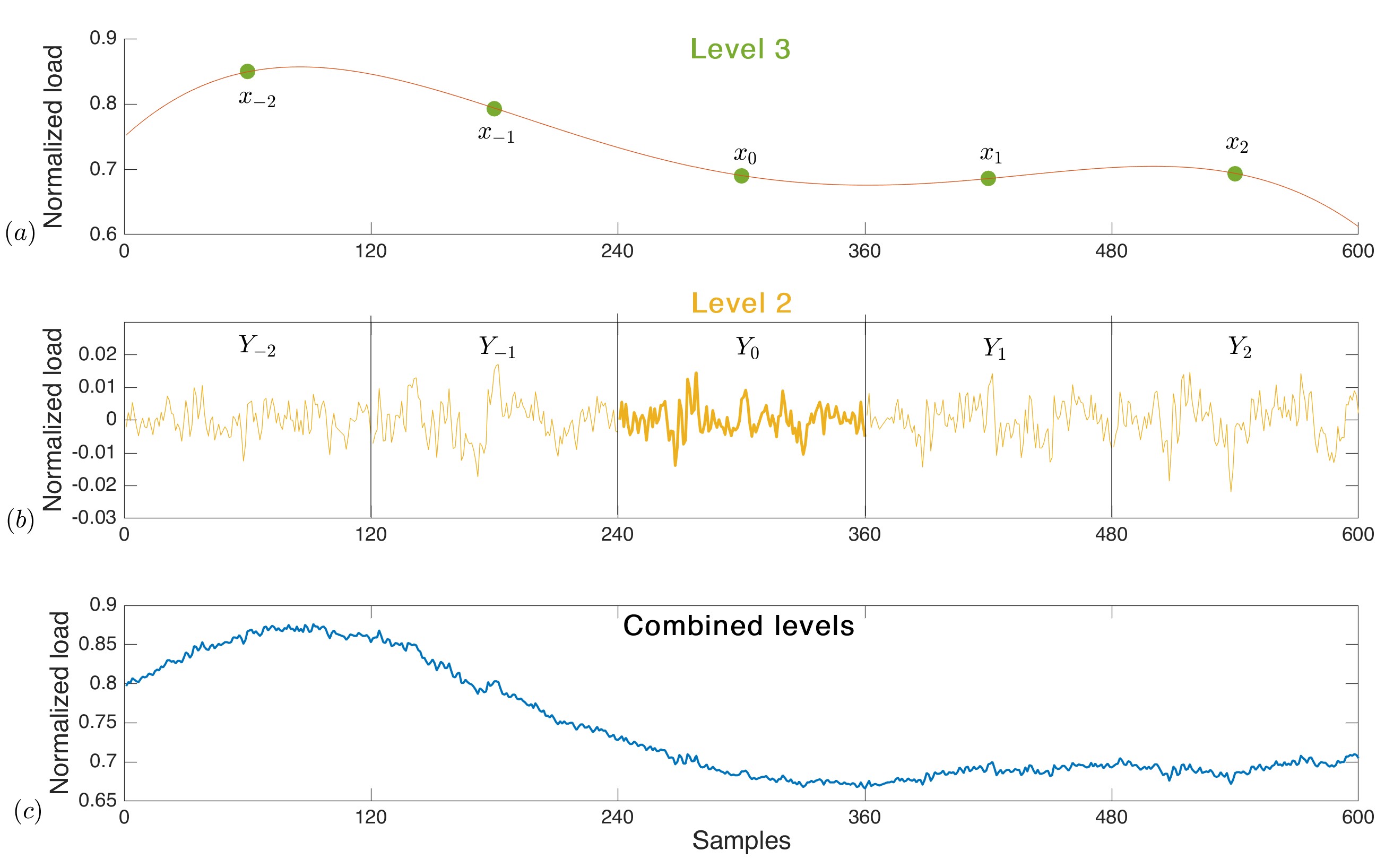}
	\caption{Illustration of the process to combine profiles from Level 3 and 2. Plot a) shows 5 hours of data at Level 3, where each point represents the average hourly load. The orange curve represents the fitted 4-degree polynomial for hour number 0 which is obtained by fitting the data representing 2 hours before and two hours after the hour of interest (hour 0 in this case). Plot b) shows 5 hour-long profiles which have been concatenated to obtain 5 hours worth of data a 1 sample every 30 seconds. Plot c) shows the result of combining the profiles together.}
	\label{hour_combine} 
\end{figure} 

\begin{figure}
	\centering
	\includegraphics[scale=0.3]{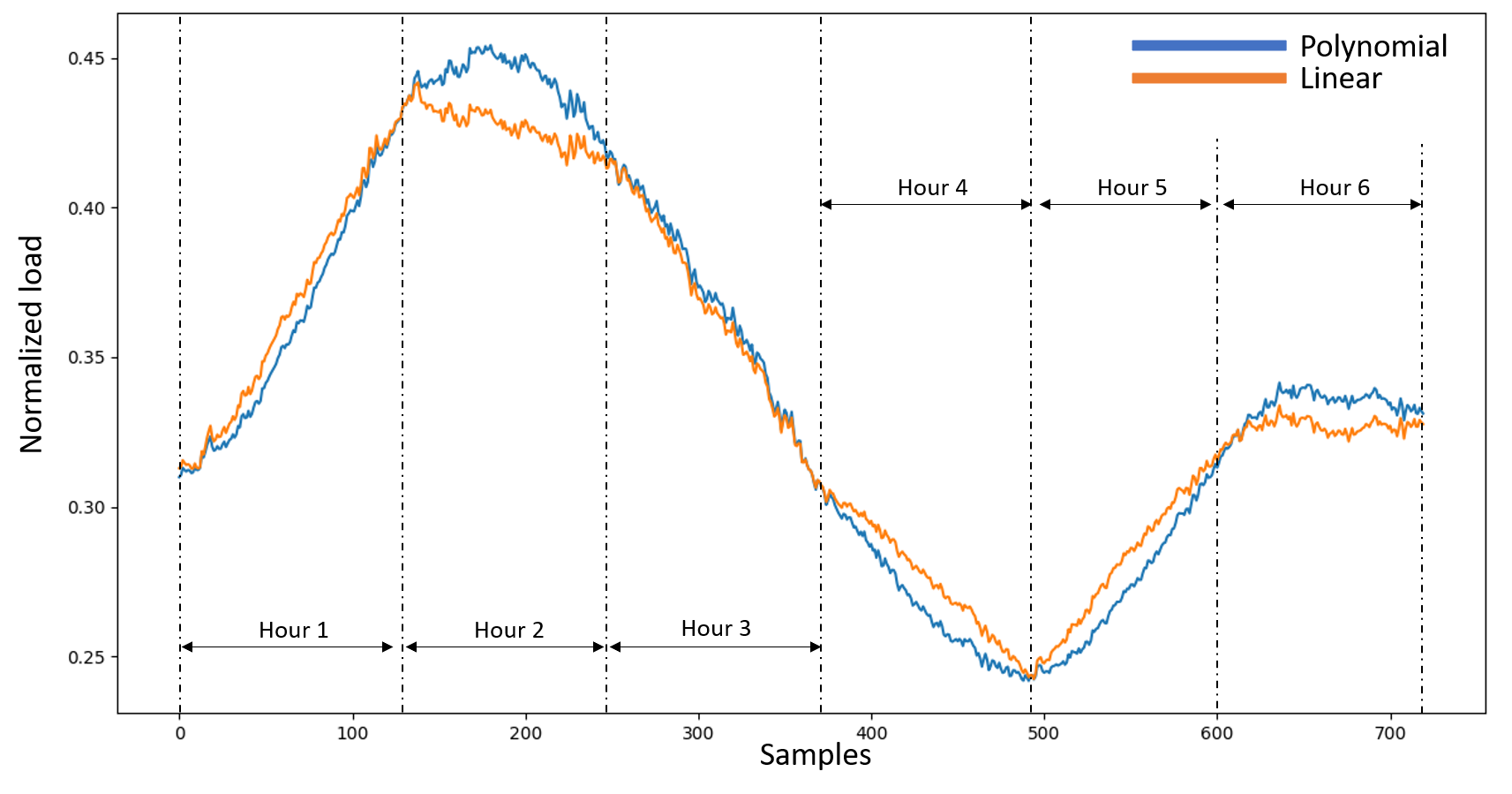}
	\caption[Combining Levels 3 and 2: Comparison Between Polynomial and Linear Approach.]{Combining Levels 3 and 2: comparison between polynomial and linear approach.}
	\label{hour_combine_example} 
\end{figure}

This particular process is illustrated in Fig.~\ref{hour_combine}: in this example, 5 hours of data are generated by combining 5 hours from Level 3 and 5 hour-long profiles from Level 2. Since each point in Fig.~\ref{hour_combine}(a) represents the average hourly load, each of the 5 hour-long profiles are first scaled based on the corresponding average values and they are then concatenated to obtain five hours of consecutive data. To capture the overall behavior observed in the Level 3 profile, the polynomial trend must be added. For example, for the hour at time 0 hour-long profile, a polynomial curve (shown in orange in Fig.~\ref{hour_combine}(a)) is fitted to the five points representing two hours before and two hours after hour 0. This trend is then added to the hour-long profile and the resulting time-series is shown in plot c). By repeating this process for each hour, the overall envelope of the load is modeled. 
At this point, it is also possible to demonstrate the validity of the polynomial trend approach by comparing it to a simpler, linear approach. Figure~\ref{hour_combine_example} shows six hours of data obtained by combining one profile from Level 3 and six profiles from Level 2. The blue curve represents the data combined using the polynomial approach, while in orange is the result of combining the profiles using a linear trend computed between two consecutive hours. It can be seen that using a $4^{th}$ degree polynomial yields a very smooth and natural behavior; on the other hand, the linear approach results in very sharp load changes where different hour-long profiles are concatenated together. 

(iii) \textit{Combining Levels 2 and 1.} The process of combining these two levels is the same as in case (i). The 30-second-long profiles of Level 1 were originally obtained by normalizing the real data by dividing them by their average value; at the same time, each point in the Level 2 profiles represents the average load over a 30 seconds period. Thus, the generated 30-second-long profiles are scaled by the corresponding value in the hour-long profiles to which they are combined with. 

\subsection{Generating load profiles at arbitrary resolutions}\label{anyres}
To make the multi-resolution generative scheme we designed applicable to the largest number of use cases and applications, a user should not be limited to the four predetermined sampling resolutions which were used to disaggregate the data. To obtain a specific resolution, first, a synthetic load profile is generated by aggregating profiles from different levels as described in Section~\ref{vercomb} in such a way that the resulting time-series will have a resolution which is higher than that requested; then, the generated data is down-sampled to the exact resolution necessary. For example, a common sampling rate for load measurements is 1 sample/10 minutes as specified by the standard IEC 61000-4-30 \cite{iec61000-4-30}. As this resolution falls between that of the week-long profiles (1 sample/hour) and hour-long profiles (1 sample/30 seconds), the synthetic data will be generated by aggregating profiles down to sampling rate of Level 2 and the resulting output is then down-sampled to achieve the correct resolution.
The down-sampling is achieved by computing either the average, minimum, or maximum load over the desired sampling period. The choice of down-sampling metric depends on the specific resolution needed as well as the user requirements. For example, if daily load values are needed, the synthetic data will be generated at the week-long level at 1 sample/hour and then the values over a 24 hour period will be aggregated. Depending on the analysis for which the data is needed, the maximum daily value could be considered in order to provide a worst case scenario or the mean or minimum values could be computed to represent lighter load conditions. 

\section{Practical Implementation}\label{practical}

\begin{figure}
	\centering
	\includegraphics[scale=0.35]{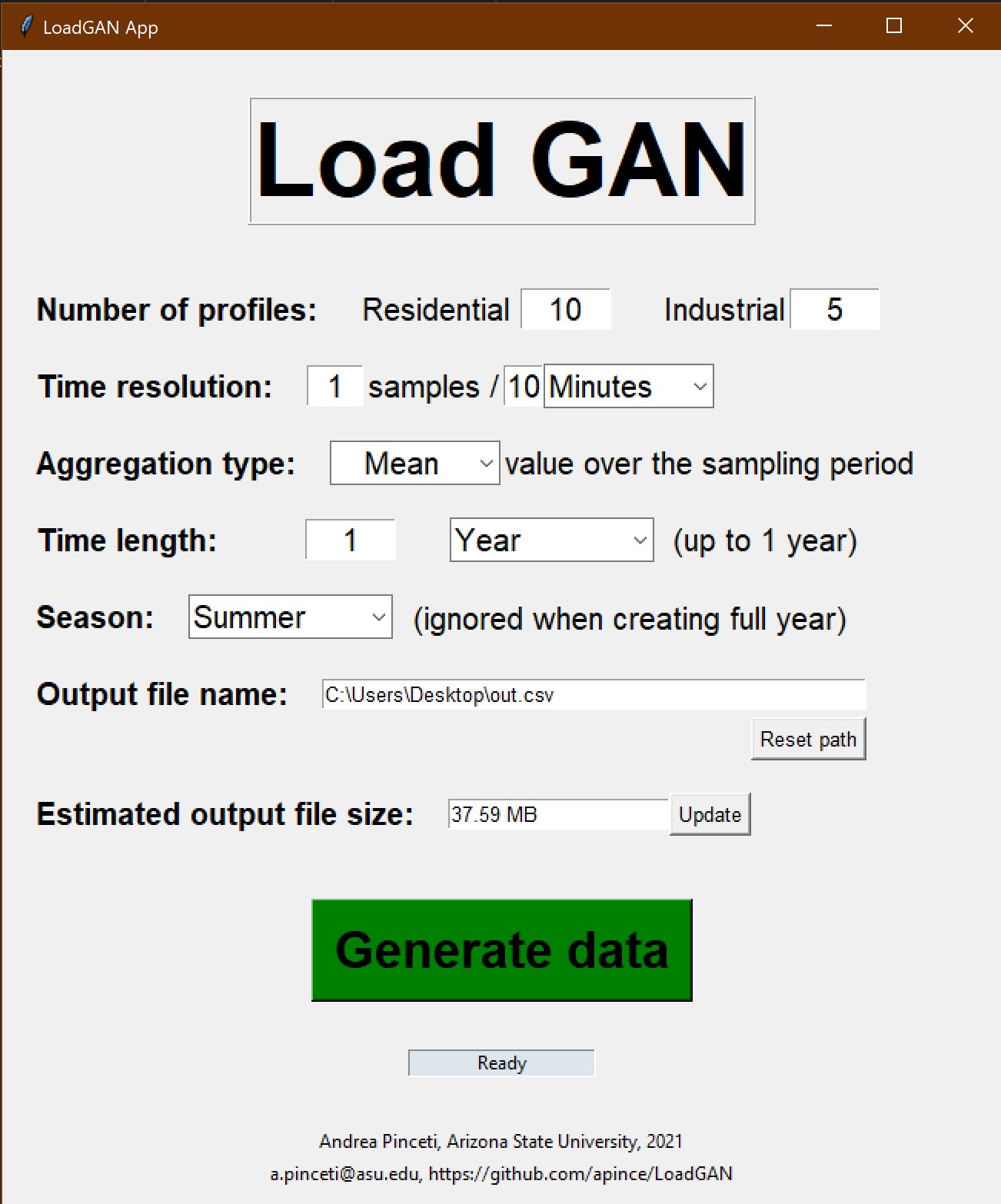}
	\caption[Interface of the LoadGAN Application.]{Interface of the LoadGAN application.}
	\label{gui} 
\end{figure} 

\subsection{LoadGAN - Open source application}\label{ch7_gui}
All the code developed as part of this project is publicly available online at \cite{appendix}. This GitHub repository contains the code used to create and train the generative models for each of the levels as well as the fully trained models which can be directly used to generate new synthetic load data. In addition to the raw code, we created an application for the generation of multi-resolution time-series data, called LoadGAN, which can be used via a graphical user interface (GUI). 

LoadGAN allows a user to generate unique synthetic data tailored to a specific application by leveraging all of the features of the generative scheme. Figure~\ref{gui} shows the settings which are available for the targeted generation of synthetic load data. Multiple profiles can be generated simultaneously by specifying the number of mainly residential and mainly industrial loads in a system. The sampling resolution of the synthetic data is entered as number of samples over sampling period; as an example, Fig.~\ref{gui} shows a resolution of 1 sample/10 minutes, as specified by IEC 61000-4-30. As described in the previous section, the aggregation to a specific resolution can be achieved by computing the mean, minimum or maximum value over the sampling period; the user can select the desired aggregation type. The temporal length of the generated data can be specified from a few seconds to one entire year. Finally, the user can select the season to be used as label for the generation of the week-long profiles. When creating a full year of data, this setting will be overridden and the data will be generated based on the correct sequence of seasons, starting in winter (January 1st). The synthetic data generated is saved in .csv format and the destination folder and file name can be modified by the user. The application also computes the estimated file size based on the currently selected settings before the data is generated.

\subsection{Validation of synthetic data}\label{validation}
In Section~\ref{loadmodelling}, the generative models for each level have been validated by performing several tests on the synthetic data generated by them. These tests showed that the synthetic load profiles retain the characteristics of the real data and they can successfully be used for power system applications. The last step in validating the compete multi-resolution generative scheme consists in verifying that it can correctly create realistic load data at any resolution. By testing synthetic data at a sampling rate different from any of the four that were initially chosen for the four aggregation levels, we can effectively validate the techniques used for the aggregation of profiles from different levels as well as the re-sampling process to obtain a specific resolution. 

The validation procedure consists in training a load forecasting algorithm solely on synthetically generated data and then testing it on the real data, similarly to what described in \cite{Pinceti2021PESGM} for the week-long profiles. Moreover, we use this form of testing to validate the claim that linear models (such as SVD) are not sufficient to adequately capture the richness of our dataset. This is done by training the forecasting algorithm on GAN-generated data and on SVD-generated data, independently, and then comparing the prediction performance. In this case, we test time-series load data sampled at 1 sample/10 minutes, which is not one of the original sampling rates used to define the Levels in the LoadGAN. A long-short term memory (LSTM) network with four layers and 36 units per layer is trained to predict the load value at a given point in time, based on the previous six hours of data (that is, the previous 36 samples). The algorithm is trained twice: first, on a dataset of about 8000 six-hours-long profiles synthetically generated using the LoadGAN application, and then on a dataset of the same size, generated using the SVD approach. The trained model is then tested on 1000 new synthetic GAN and SVD profiles as well as 1000 profiles from real load data; to evaluate the performance, the absolute percentage difference between the actual and predicted value is computed. Table~\ref{ch7_forecasttable} shows the summary statistics of the prediction error of the two load forecasters. These results allow for two main observations. First, it can be noted that when training the LSTM on GAN synthetic data, the error across the three testing datasets is very similar. This means that although the model was only trained on synthetic data, it effectively captures the behavior of the real data. In addition to this, the model trained on GAN data performs very well when tested on SVD data; however, the model trained on SVD data shows an error which is about twice as large when tested on GAN and real data. Based on these results, it can be confirmed that the generative model based on the linear approach is not able to fully represent the wide range of behaviors of real and GAN-generated load data.

\begin{table}
\centering
\caption{Comparison of the forecasting error between generated and real load data, sampled at 1 sample/10 minutes.}
\label{ch7_forecasttable}
\begin{tabular}{|c|c|c|c|} 
\hline
\multicolumn{1}{|l|}{\multirow{2}{*}{\textbf{Training Dataset}}} & \multicolumn{1}{l|}{\multirow{2}{*}{\textbf{Testing Dataset}}} & \multicolumn{2}{c|}{\textbf{Percentage Error}} \\ 
\cline{3-4}
\multicolumn{1}{|l|}{} & \multicolumn{1}{l|}{} & ~ ~Mean~ ~ & \begin{tabular}[c]{@{}c@{}}Std. Dev.\\\end{tabular} \\ 
\hline
\hline
\multirow{3}{*}{GAN} & GAN & 2.22 & 2.16 \\ 
\cline{2-4}
 & SVD & 2.49 & 1.86 \\ 
\cline{2-4}
 & Real & 2.28 & 2.90 \\ 
\hline
\hline
\multirow{3}{*}{SVD} & SVD & 1.86 & 2.66 \\ 
\cline{2-4}
 & GAN & 3.71 & 2.93 \\ 
\cline{2-4}
 & Real & 3.63 & 2.92 \\
\hline
\end{tabular}
\end{table}

\section{Conclusion}

A complete scheme for the generation of time-series load data is proposed. To capture load behaviors at different time-scales, the generative framework leverages multiple models (GANs, cGANs, and PCA-based models). By appropriately combining these models, realistic load data of any time length and resolution can be generated. Each of these models is extensively tested and the aggregation scheme is carefully designed to produce high-quality, realistic time-series data. All of these components have been compiled into a simple open-source graphical interface that, along with the trained generative models, allows an end-user to quickly generate any amount of synthetic load data based on selectable parameters. 

More broadly, the generative scheme we designed represents a flexible framework which can be tuned to capture an even larger number of interesting load characteristics. For example, the generation of representative load scenarios is crucial to ensure the reliability of our fast-evolving electrical grid. The ability of the proposed generative scheme to learn and subsequently generate data conditioned on different load characteristics could be leveraged for the efficient modeling of loads with varying degrees of penetration of electrical vehicles and/or distributed energy resources.


\bibliography{LoadGAN}
\bibliographystyle{ieeetran}

\end{document}